\documentclass[conference]{IEEEtran}
\IEEEoverridecommandlockouts
\usepackage{cite}
\usepackage{amsmath,amssymb,amsfonts}
\usepackage{algorithmic}
\usepackage{graphicx}
\usepackage{textcomp}
\usepackage{xcolor}
\usepackage{subcaption}

\def\BibTeX{{\rm B\kern-.05em{\sc i\kern-.025em b}\kern-.08em
    T\kern-.1667em\lower.7ex\hbox{E}\kern-.125emX}}
    
\begin{document}

\title{3-Lead to 12-Lead ECG Reconstruction: A Novel AI-based Spatio-Temporal Method\\
}

\author{\IEEEauthorblockN{Rahul LR}
\IEEEauthorblockA{\textit{Dept. of Electrical Engineering} \\
\textit{IIT Hyderabad}\\
Hyderabad, India \\
ee18resch01004@iith.ac.in}
\and
\IEEEauthorblockN{Albert Shaiju}
\IEEEauthorblockA{\textit{Dept. of Electrical Engineering} \\
\textit{NIT Calicut}\\
Calicut, India \\
albertshaju100@gmail.com}
\and
\IEEEauthorblockN{Soumya Jana}
\IEEEauthorblockA{\textit{Dept. of Electrical Engineering} \\
\textit{IIT Hyderabad}\\
Hyderabad, India \\
jana@ee.iith.ac.in}

}

\maketitle

\begin{abstract}

Diagnosis of cardiovascular diseases usually relies on the widely used standard 12-Lead (S12) ECG system. However, such a system could be bulky, too resource-intensive, and too specialized for personalized home-based monitoring. In contrast, clinicians are generally not trained on the alternative proposal, i.e., the reduced lead (RL) system. This necessitates mapping RL to S12. In this context, to improve upon traditional linear transformation (LT) techniques, artificial intelligence (AI) approaches like long short-term memory (LSTM) networks capturing non-linear temporal dependencies, have been suggested. However, LSTM does not adequately interpolate spatially (in 3D). To fill this gap, we propose a combined LSTM-UNet model that also handles spatial aspects of the problem, and demonstrate performance improvement.
Evaluated on PhysioNet PTBDB database, our LSTM-UNet achieved a mean $R^2$ value of 94.37\%, surpassing LSTM by 0.79\% and LT by 2.73\%. Similarly, for PhysioNet INCARTDB database, LSTM-UNet achieved a mean $R^2$ value of 93.91\%, outperforming LSTM by 1.78\% and LT by 12.17\%.

\end{abstract}

\begin{IEEEkeywords}
Artificial Intelligence, Electrocardiogram, LSTM, ECG Reconstruction, Reduced 3-Lead ECG, Standard 12-Lead ECG, UNet.
\end{IEEEkeywords}


\begin{figure}

\centering
\begin{subfigure}{0.22\textwidth}
\includegraphics[width=\textwidth]{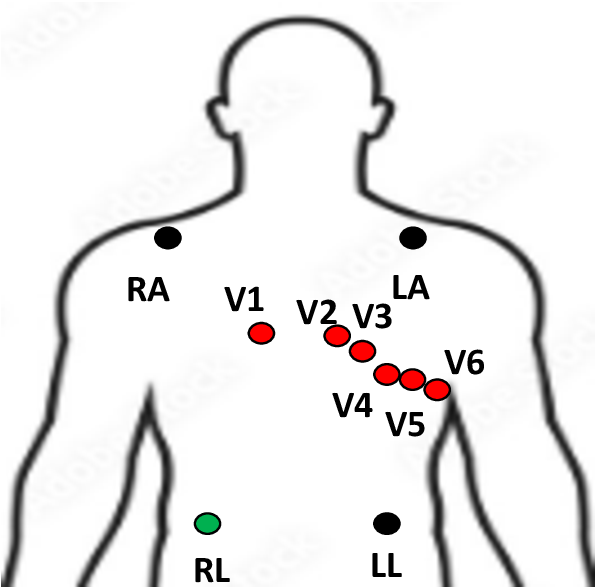}
\caption{}
\label{fig:figure1}
\end{subfigure}
\quad
\begin{subfigure}{0.22\textwidth}
\includegraphics[width=\textwidth]{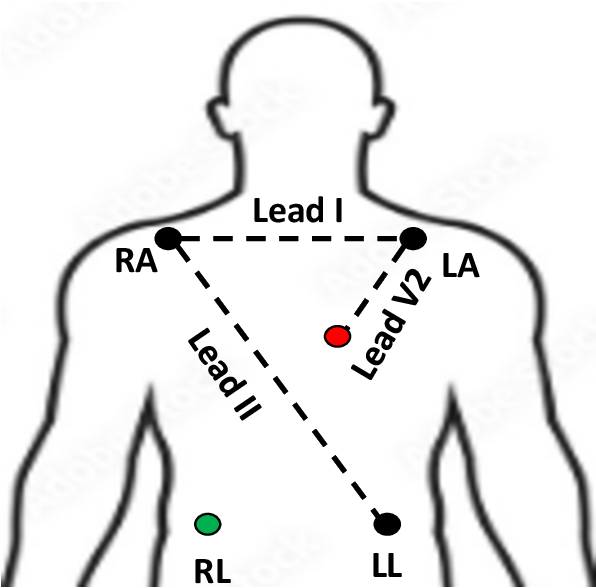}
\caption{}
\label{fig:figure2}
\end{subfigure}

  
\begin{subfigure}{0.5\textwidth}
\centering
\includegraphics[width=0.6\textwidth]{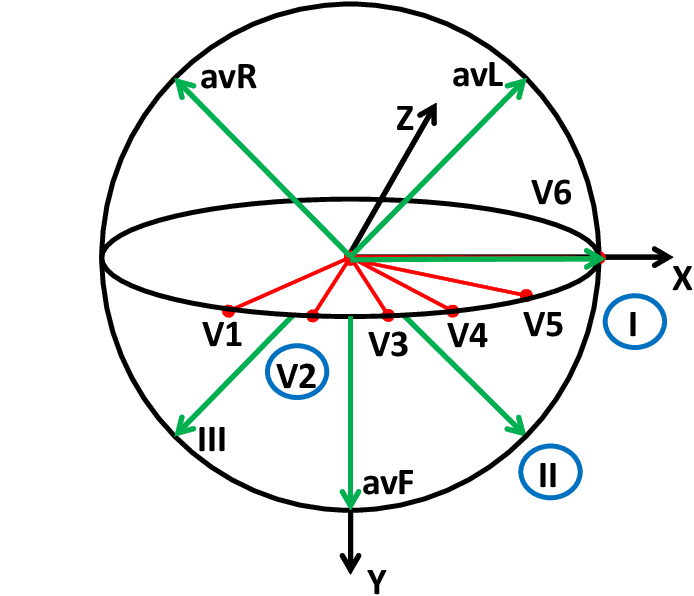}
\caption{}
\label{fig:figure3}
\end{subfigure}

\caption{(a) Standard 12-Lead ECG system; (b) Reduced 3-Lead ECG system; (c) Schematic presentation of the lead vectors of Standard 12-lead ECG system with the 3 leads in the reduced system circled.}
\label{S12_R3L_3D}

\end{figure}

\begin{figure*}
\includegraphics[width=0.9\textwidth]{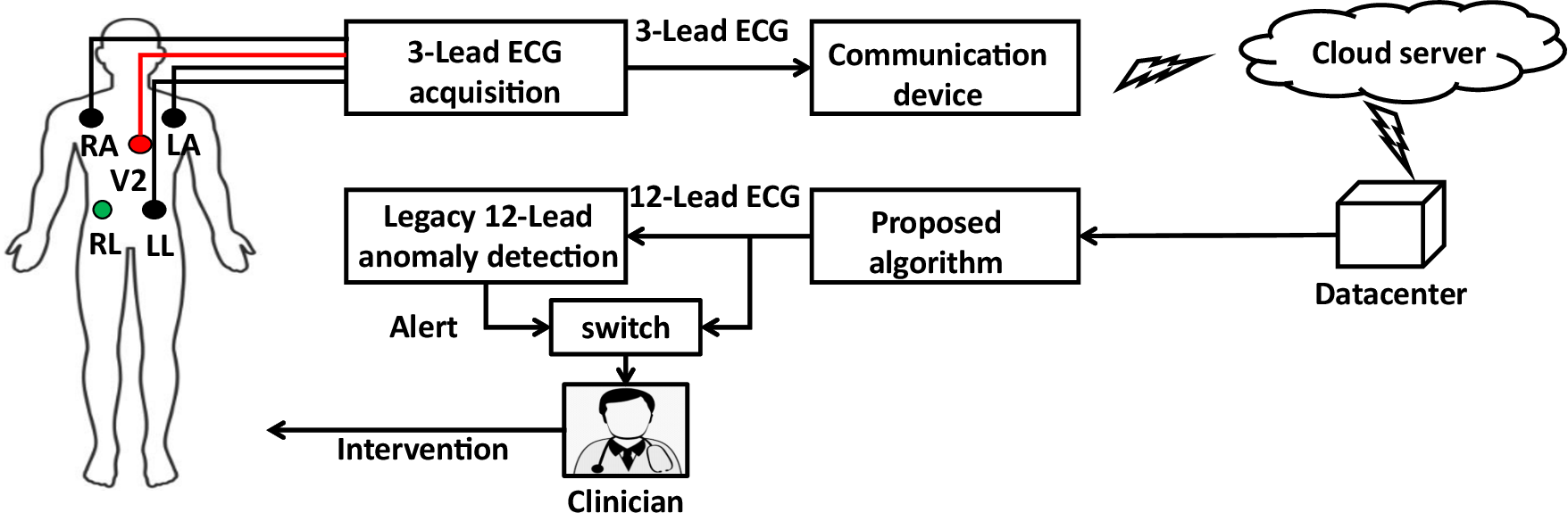}
\centering
\caption{Block schematic of an envisaged personalized home-based cardiac monitoring system.}
\label{home_monitoring}
\end{figure*}

\section{Introduction}


The Electrocardiogram (ECG), obtained using the Standard 12-Lead (S12) system, is a widely utilized medical technology for diagnosing cardiovascular diseases (CVDs) \cite{12lead}. While doctors are extensively trained in diagnosing using the S12 ECG system, there has been a growing interest in the use of Reduced Lead (RL) ECG systems due to their simplicity and ease of use \cite{Rlead}. However, the RL systems poses challenges in capturing the complete electrical activity of the heart, potentially leading to diagnostic limitations. On the other hand, the S12 system has its own drawbacks, such as the requirement for multiple electrodes (see Figure \ref{S12_R3L_3D}(a)), increased complexity, limited feasibility in personalized settings and significant storage and bandwidth demands in telemonitoring applications. To address these limitations, researchers have proposed various approaches of reconstructing S12 from RL systems. In general, Lead I, II, and V2, collectively known as the reduced 3 leads (R3L) (Figure \ref{S12_R3L_3D}(b)) are commonly selected as input leads due to their ability to provide a good representation of different regions of the heart and have demonstrated diagnostic value in many cardiac conditions \cite{R3lead}. Thus, with 3-Lead to 12-Lead reconstruction, we bridge the gap between the limited lead information of R3L and the comprehensive 3D information (Figure \ref{S12_R3L_3D}(c)) of the S12 system. This reconstruction technique enables healthcare professionals to make accurate diagnoses even in situations where only a limited number of leads are available. 


In this paper, we envisage a personalized home-based cardiac monitoring system, as depicted in Figure \ref{home_monitoring}. The proposed algorithm reconstructs the missing leads of S12 system from the R3L system. However, this reconstruction process poses a significant engineering challenge, as it requires accurate estimation of the missing information to create a comprehensive and reliable representation of the cardiac electrical activity. This challenge can be attributed to several factors that impact the accuracy and precision of the reconstruction process.
First, the choice between reconstructing using orthogonal leads (eg. Frank vectorcardiographic system) or a subset of the 12 leads (known as reduced lead set). Eventhough, both approaches have their merits and considerations, many recent works have focused on using a subset of the 12-lead ECG system rather than orthogonal leads due to practical considerations, diagnostic relevance, computational efficiency and allows for better validation and comparison.  
Second, the selection of leads plays a critical role in capturing essential cardiac information. Each lead provides unique spatial information about the electrical activity of the heart, and therefore identifying the optimal combination of leads becomes crucial to ensure accurate reconstruction. Usually, Lead I, II, and V2 were chosen due to their diagnostic significance and their ability to provide valuable information about the heart's electrical activity from different perspectives. 
Moreover, the reconstruction process needs to address the temporal dynamics of the cardiac electrical activity. The heart's electrical signals change over time, and capturing these dynamic patterns is crucial for an accurate reconstruction. Devising effective algorithms that can accurately model and predict these spatio-temporal changes in ECG is an ongoing engineering challenge.



\begin{figure*}

\centering
\begin{subfigure}{0.48\textwidth}
\includegraphics[width=\textwidth]{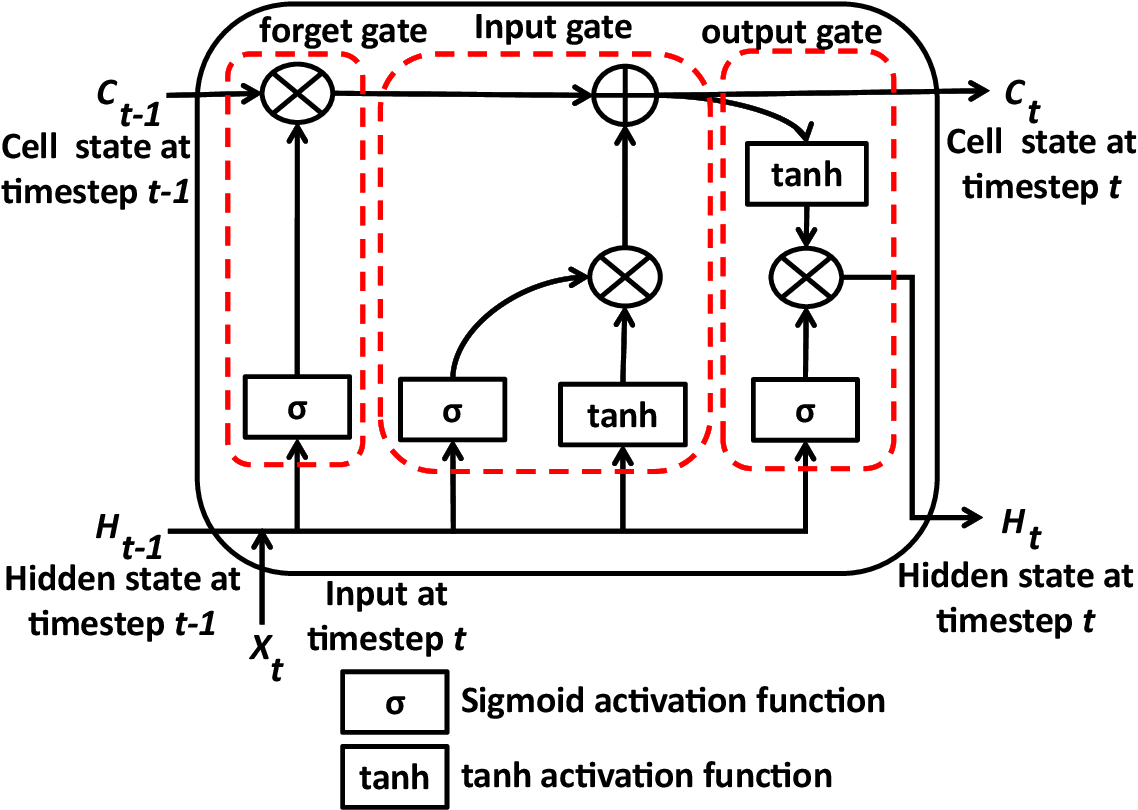}
\caption{}
\label{LSTM}
\end{subfigure}
\quad
\begin{subfigure}{0.48\textwidth}
\includegraphics[width=\textwidth]{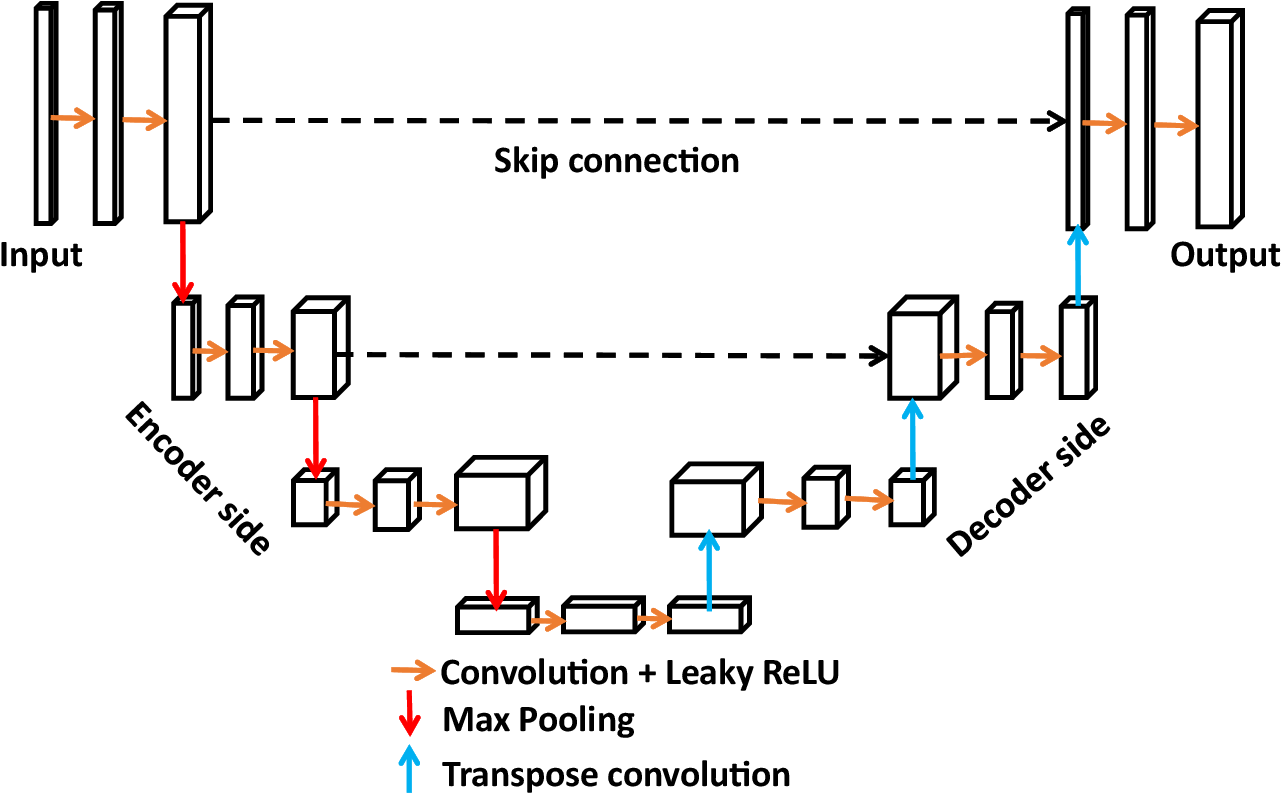}
\caption{}
\label{UNet}
\end{subfigure}

  
\begin{subfigure}{0.95\textwidth}
\centering
\includegraphics[width=\textwidth]{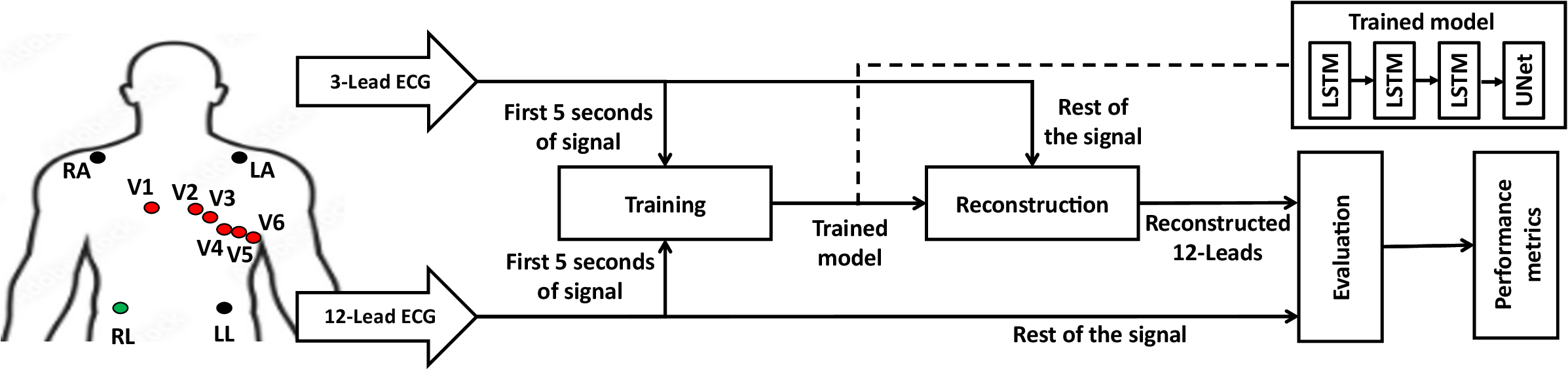}
\caption{}
\label{block_diagram}
\end{subfigure}

\caption{(a) LSTM unit; (b) UNet architecture; (c) 3-Lead to 12-Lead reconstruction: Proposed LSTM-UNet method.}
\label{LSTM_UNET_block}

\end{figure*}

To overcome the aforementioned challenges, engineers employ various techniques which includes linear transformation (LT) methods, statistical modelling techniques and machine learning (ML) approaches. 
In prior research, Dower introduced the Dower Transform (DT) to convert the Frank vectorcardiographic system to the S12 system \cite{dower}. Subsequently, an improved linear Affine transformation method using the Least-Square (LS) fit approach was proposed, demonstrating superior performance over DT \cite{affine}. 
\cite{amit} used a linear transformation (LT) method to map R3L to S12 system where a Least square fit and heart vector projection theory have been used to obtain personalized transformation coefficients. 
Even though the linear transformation methods yield acceptable reconstruction, recent studies have shown that ML based methods can further improve the quality of the reconstructed ECG \cite{ann}. \cite{lstm} designed a hardware for a 3-Lead patch type device and employed a Long short-term memory (LSTM) network for missing lead reconstruction. This network can capture the non-linear temporal dependencies between the leads which LT methods struggle to capture. However, the LSTM still have limitations in capturing spatial dependencies and local features. 


Against this backdrop, we note studies indicating the capability of UNet architecture to operate at multiscale, enabling them to capture spatial dependencies \cite{unet}. Accordingly, we propose to develop a LSTM-UNet model, which effectively captures and integrates spatio-temporal features for ECG reconstruction. When evaluated on publicly available datasets, the proposed algorithm demonstrates superior performance compared to existing state-of-the-art methods.

\section{Materials and Methods}

The task of 3-Lead to 12-Lead reconstruction was posed as sequence-to-sequence translation problem to be solved using a LSTM-UNet model. 

\subsection{Database}

The PhysioNet’s PTB Database (PTBDB) and St. Petersburg Institute of Cardiological Technics 12-Lead Arrhythmia Database (INCARTDB) have been used in this study \cite{physionet}.  
Patients in PTBDB were classified into two groups: Healthy control (HC) and diseased. The diseased group was further subdivided into 5 categories: bundle branch block (BB), hypertrophy, cardiomyopathy, and heart failure (HY), myocardial infarction (MI), valvular myocarditis, and other miscellaneous (VA), and patients with no diagnostic data (ND). Conversely, the INCARTDB has records of wide range of heart conditions, but did not have separate categories. The accuracy of reconstruction was evaluated by comparing the performance of LSTM-UNet with that of LT \cite{amit} and LSTM \cite{lstm} on the Physionet PTBDB and INCARTDB.  

\subsection{Linear Transformation (LT)}

Linear transformation (LT) methods for ECG reconstruction involve using mathematical transformations to map a R3L system to the S12 system. The personalized transformation coefficients can be obtained using the Heart Vector Projection (HVP) theory and Least Square (LS) fit method \cite{amit}. The HVP theory allows for the estimation of transformation vectors that project the R3L leads onto the corresponding S12 leads. The LS fit method optimizes these transformation vectors to minimize the error between the original and transformed ECG signals, resulting in accurate reconstruction of the missing leads. Personalized coefficients enhance accuracy by tailoring ECG reconstruction to individual signals, surpassing traditional linear methods.


\begin{figure*}

\centering
\begin{subfigure}{0.48\textwidth}
\includegraphics[width=\textwidth,height=.6\textheight]{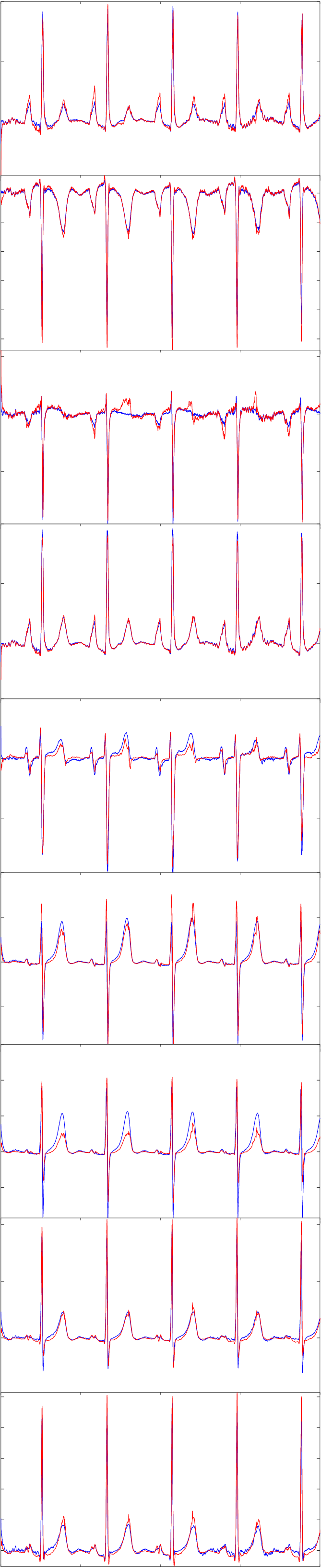}
\caption{}
\label{LSTM}
\end{subfigure}
\quad
\begin{subfigure}{0.48\textwidth}
\includegraphics[width=\textwidth,height=.6\textheight]{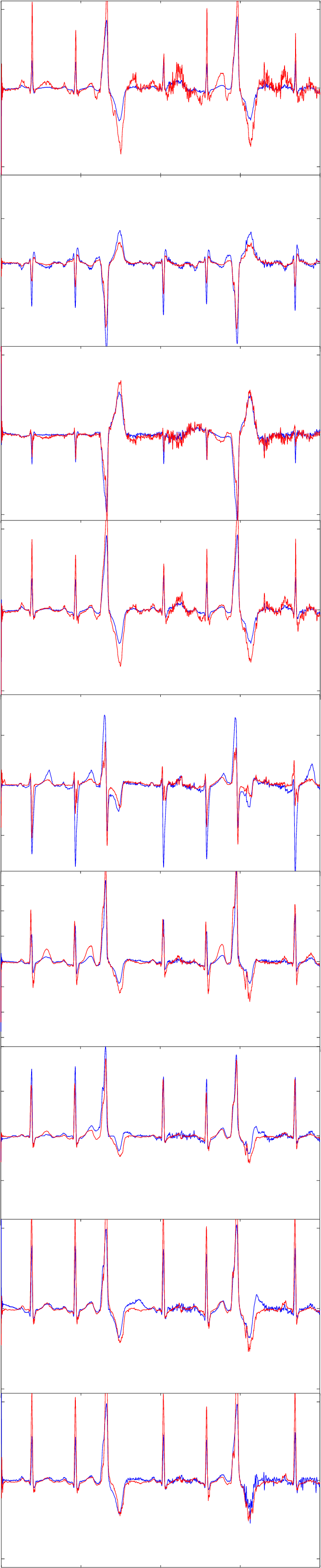}
\caption{}
\label{UNet}
\end{subfigure}

\caption{From leads I, II and V2, reconstruction of leads III, aVR, aVL, aVF, V1, V3, V4, V5, V6 illustrated in each subfigure (in that order, top to bottom) belonging to representative (a) healthy subject and (b) unhealthy subject from the PTBDB.}
\label{HC_UH}

\end{figure*}

\begin{table*}[t!]
\begin{center}

\caption{Average $R^2$, $r_x$, and $NDTW$ between the original and reconstructed ECG leads/signals for different patient groups in PTBDB: column `a' refers to the reference Linear Transformation (LT) method \cite{amit}, `b' to the reference LSTM \cite{lstm}, and `c' to the proposed LSTM-UNet model. For column `a', average $R^2$ and $r_x$ values are available only for precordial leads (V1-V6) \cite{amit}. Each of Subtables (a)-(f) pertains to a patient group exhibiting a specific type of heart condition:
(a) Health control (HC), (b) bundle branch block (BB), (c) hypertrophy, cardiomyopathy, and heart failure (HY), (d) myocardial infarction (MI), (e) valvular myocarditis, and other miscellaneous (VA), and (f) patients with no diagnostic data (ND).
}
\renewcommand{\arraystretch}{1.1}
\renewcommand{\tabcolsep}{3pt}
\begin{tabular}{cc}
{
\begin{small}
\begin{tabular}{|c|ccc|ccc|cc|}
\hline
\cline{2-9}
& \multicolumn{3}{c|}{$R^2$}
& \multicolumn{3}{c|}{$r_x$} & \multicolumn{2}{c|}{$NDTW$} 
\\
\cline{2-9}
Lead & a & b & c & a & b & c & b & c \\
\hline
III & - &95.37 & 96.22 & -  & 96.53& 98.17 & 0.0079 & 0.0075\\
avR & - & 99.25 & 99.25 & - & 99.48 & 99.63 & 0.0031 & 0.0032\\
avL & - & 96.37 & 97.06 & -& 96.45 & 98.61 & 0.009 & 0.0068\\
avF & - & 97.97 & 98.35 & - & 98.45& 99.2 & 0.004& 0.004\\
V1 & 94.69 & 97.65 & 97.83 & 97.3 & 97.64 & 98.94& 0.0063 & 0.0093\\
V3 & 96.52 & 97.64 & 97.64 & 98.3 & 98.56& 98.87 & 0.0066 & 0.0056\\
V4 & 91.83 & 96.29 & 96.54 & 95.9 & 97.85 & 98.35 & 0.0071 & 0.0086\\
V5 & 93.53 & 95.38 & 95.63 & 96.8 & 98.26 & 97.96& 0.0061 & 0.0065\\
V6 & 94.69 & 97.15 &97.62 & 97.3 & 98.38 & 99 & 0.0077& 0.0038\\
\hline
Avg & 94.25 & 96.82 & 97.05 & 97.12 & 98.14 & 98.62 & 0.0064 & 0.0061\\

\hline
\end{tabular}
\end{small}
}
&
{
\begin{small}
\begin{tabular}{|c|ccc|ccc|cc|}
\hline
\cline{2-9}
& \multicolumn{3}{c|}{$R^2$}
& \multicolumn{3}{c|}{$r_x$} & \multicolumn{2}{c|}{$NDTW$} 
\\
\cline{2-9}
Lead & a & b & c & a & b & c & b & c \\
\hline

III & -& 90.70 & 91.32 & - & 95.79 & 96.13 & 0.0138 & 0.0141\\
avR & -& 93.58 & 93.78& - & 96.77& 96.85 & 0.0065 & 0.0080\\
avL & - & 95.97 & 95.99 & - & 98.02 & 98.02 & 0.0111 & 0.0106\\
avF & - & 96.19 & 96.44 & - & 98.21 & 98.38 & 0.0074 & 0.0081\\
V1 & 90.40 & 94.71 & 95.33 & 94.7& 97.34 & 97.67 & 0.0158 & 0.0083\\
V3 & 95.78 & 97.04 & 97.31 & 97.9& 98.56 & 98.70 & 0.0053 & 0.0083\\
V4 & 87.39 & 92.70 & 93.80& 93.7 & 96.55 & 97.00 & 0.0085 & 0.0085\\
V5 & 88.47 & 89.78 & 89.96 & 94.4 & 95.26 &95.44 &0.0066 & 0.0075\\
V6 & 92.41 & 90.12 & 90.19 & 96.2 & 95.89 & 95.94 & 0.0083 & 0.0089\\
\hline
Avg & 90.89 & 92.87& 93.32 & 95.38 & 96.72 & 96.95& 0.0093 & 0.0091\\

\hline
\end{tabular}
\end{small}
} \\
(a) & (b)\\
{
\begin{small}
\begin{tabular}{|c|ccc|ccc|cc|}
\hline
\cline{2-9}
& \multicolumn{3}{c|}{$R^2$}
& \multicolumn{3}{c|}{$r_x$} & \multicolumn{2}{c|}{$NDTW$} 
\\
\cline{2-9}
Lead & a & b & c & a & b & c & b & c \\
\hline
III & - & 95.51 & 96.11 & - & 97.88 & 98.10 & 0.0098 & 0.0089\\
avR & - & 97.49 & 98.01 & - & 92.78 & 93.60 & 0.0072 & 0.0057\\
avL & - & 95.66 & 96.58 & - & 97.93& 98.30& 0.0082 & 0.0097\\
avF & - & 96.98 & 97.42 & - & 98.56 & 98.74 & 0.0098 & 0.0058\\
V1 & 97.44 & 96.81& 97.12 & 98.7 & 98.43 &98.59 & 0.0076 & 0.0073\\
V3 & 96.23 & 97.09 & 97.87 & 98.1 & 98.61 & 98.98& 0.0066 & 0.0085\\
V4 & 87.45 & 92.41& 94.07& 93.4 & 96.40 & 97.11 & 0.0119 & 0.0090\\
V5 & 89.90 & 91.64 & 92.41 & 94.5 & 96.04 & 96.44& 0.0094 & 0.0076\\
V6 & 95.74 & 93.65 & 94.54 & 97.8 & 96.98 & 97.47 & 0.0078 & 0.0075\\
\hline
Avg & 93.35 & 94.32 & 95.20 & 96.5 & 97.29 & 97.72& 0.0087 & 0.0078\\
\hline
\end{tabular}
\end{small}
}
&
{
\begin{small}
\begin{tabular}{|c|ccc|ccc|cc|}
\hline
\cline{2-9}
& \multicolumn{3}{c|}{$R^2$}
& \multicolumn{3}{c|}{$r_x$} & \multicolumn{2}{c|}{$NDTW$} 
\\
\cline{2-9}
Lead & a & b & c & a & b & c & b & c \\
\hline
III & - & 91.45 & 91.61 & - & 95.97& 96.06 & 0.0124 & 0.0122\\
avR & - & 97.39 & 97.46 & - & 98.70 & 98.73 & 0.0069 & 0.006\\
avL & - & 95.01 & 95.04 & - & 97.38 & 97.86 & 0.0078 & 0.008\\
avF & - & 94.19 & 94.22 & - & 97.18 & 97.25 & 0.0076& 0.0073\\
V1 & 94.34& 93.72 & 94.60 & 97.1 & 96.90 & 97.34 & 0.0117 & 0.0103\\
V3 & 95.92 & 95.48 & 95.87 & 97.9 & 97.80 & 98.00 & 0.0079& 0.0077\\
V4 & 89.52 & 89.75 & 90.44 & 95.4 & 95.55 & 95.57& 0.0108 & 0.0099\\
V5 & 89.22 & 90.12 & 90.20 & 94.4 &95.16 & 95.44 & 0.0096 & 0.0077\\
V6 & 91.52 & 92.53 & 92.97 & 95.6 & 95.66 & 95.83 & 0.0091 & 0.013\\
\hline
Avg & 92.10 & 92.32 & 92.82 & 96.08 & 96.21 & 96.44 & 0.0093 & 0.0091\\
\hline
\end{tabular}
\end{small}
}
\\
(c) & (d)\\
{
\begin{small}
\begin{tabular}{|c|ccc|ccc|cc|}
\hline
\cline{2-9}
& \multicolumn{3}{c|}{$R^2$}
& \multicolumn{3}{c|}{$r_x$} & \multicolumn{2}{c|}{$NDTW$} 
\\
\cline{2-9}
Lead & a & b & c & a & b & c & b & c \\
\hline
III & - & 92.67 & 92.85 & - & 95.74 & 96.55 & 0.0089 & 0.0072\\
avR & - & 98.96 & 99.01 & - & 99.62 & 99.5 & 0.0047 & 0.0043\\
avL & - &93.09 & 93.98 & - & 98.21 & 97.07 & 0.0073& 0.0076\\
avF & - & 96.04 & 96.95 & - & 99.00& 98.53 & 0.0048 & 0.0051\\
V1 & 93.73 & 95.17 & 95.87 & 96.9& 98.83& 97.97 & 0.0077& 0.0081\\
V3 & 94.25 & 97.04& 97.53 & 97.0 & 98.68 & 98.83 & 0.0081& 0.0079\\
V4 & 89.97 & 93.47 & 95.46 & 94.8 & 98.19& 98.36& 0.0088 & 0.0082\\
V5 & 91.89 & 96.5 & 97.04 & 95.9 & 96.06& 98.6 & 0.006 & 0.0064\\
V6 & 93.78& 96.75 & 97.88 & 96.8 & 98.68 & 98.57 & 0.0074 & 0.0069\\
\hline
Avg & 92.72 & 95.79 & 96.56 & 96.28 & 98.09 & 98.47 & 0.0071 & 0.0069\\
\hline
\end{tabular}
\end{small}
}
&
{
\begin{small}
\begin{tabular}{|c|ccc|ccc|cc|}
\hline
\cline{2-9}
& \multicolumn{3}{c|}{$R^2$}
& \multicolumn{3}{c|}{$r_x$} & \multicolumn{2}{c|}{$NDTW$} 
\\
\cline{2-9}
Lead & a & b & c & a & b & c & b & c \\
\hline
III & - & 90.19 & 90.33 & - & 95.05 & 95.61 & 0.0118 & 0.0115\\
avR & - & 95.43 & 95.73 & - & 98.25 & 97.92 & 0.0067 & 0.0068\\
avL & - & 94.99 & 95.26 & - & 97.50 & 97.70 & 0.0095 & 0.0095\\
avF & - & 95.05 & 95.37 & - &97.56 & 97.79 & 0.0087 & 0.008\\
V1 & 90.92 & 92.69 & 93.06 & 94.9& 96.36 & 96.60 & 0.0097 & 0.0091\\
V3 & 93.54 & 94.02 & 94.51 & 96.7 & 96.47 & 97.35 & 0.0093 & 0.0099\\
V4 & 83.72 & 89.05& 89.62 & 91.0 & 94.96 & 95.10 & 0.0128 & 0.0119\\
V5 & 84.61 &86.79& 91.14 & 91.6 & 94.03 & 95.78 & 0.0108 & 0.0104\\
V6 & 87.28 & 85.66 & 87.85 & 92.0 & 93.42 & 94.35& 0.0099 & 0.0095\\
\hline
Avg & 88.01 & 89.64 & 91.24 & 93.24 & 95.05 & 95.84 & 0.0099& 0.0096\\
\hline
\end{tabular}
\end{small}
}
\\
(e) & (f)\\
\end{tabular}
\label{ptbdb_results}

\end{center}
\end{table*}

\begin{table}
\begin{center}
\begin{small}
\renewcommand{\arraystretch}{1.1}
\renewcommand{\tabcolsep}{3pt}

\caption{Average $R^2$, $r_x$, and $NDTW$ between the original and reconstructed ECG leads/signals for all patients in INCARTDB: column `a' refers to the reference Linear Transformation (LT) method \cite{amit}, `b' to the reference LSTM \cite{lstm}, and `c' to the proposed LSTM-UNet model. For column `a', average $R^2$ and $r_x$ values are available only for precordial leads (V1-V6) \cite{amit}.}

\begin{tabular}{|c|ccc|ccc|cc|}
\hline
\cline{2-9}
& \multicolumn{3}{c|}{$R^2$}
& \multicolumn{3}{c|}{$r_x$} & \multicolumn{2}{c|}{$NDTW$} 
\\
\cline{2-9}
Lead & a & b & c & a & b & c & b & c \\
\hline
III &- & 95.09 & 96.48 & - & 97.64 & 98.25 & 0.0080 & 0.0065\\
aVR & - & 96.61 & 97.60 & - & 98.35 & 98.80 & 0.0055 &0.0052\\
aVL & -& 91.51 & 93.91 & -& 95.85 & 96.98 & 0.0091 & 0.0077 \\
aVF &-& 97.28 & 98.27 & - & 98.64 & 99.17 & 0.0055 & 0.0049\\
V1 & 86.38 & 93.23 & 94.92 & 93.6& 96.73 & 97.46 & 0.0088& 0.0063\\
V3 & 86.91 & 91.52 & 94.30 & 94.0& 96.20 & 97.18 & 0.0113 & 0.0092\\
V4 & 83.61 & 91.64 & 93.85 & 92.2 & 96.04 & 96.97 & 0.0100 & 0.0101\\
V5 & 83.74 & 92.50 & 92.88 & 92.1 & 96.36&96.41 & 0.0086& 0.0083\\
V6 & 78.11 & 92.41 & 93.61 & 89.3 & 96.24 & 96.81 & 0.0094& 0.0087\\
\hline
Avg & 83.75 & 92.26 & 93.91 & 92.24 & 96.31 & 96.97 & 0.0085 & 0.0074\\

\hline
\end{tabular}
\label{incart_results}
\end{small}

\vspace{5pt}


\end{center}
\end{table}

\subsection{Long short-term memory (LSTM)}

Traditional neural networks struggle with processing sequential data because they are designed for feedforward tasks, where the input and output are independent of each other. In contrast, sequential data, such as time series, have a temporal nature, where the current data point is influenced by previous data points. Earlier, Recurrent Neural Network (RNN) were widely used for sequential data processing due to their ability to capture temporal dependencies. However, RNNs suffer from the vanishing/exploding gradient problem, which hinders their capability to retain and propagate long-term dependencies. As a result, LSTM (Figure \ref{LSTM_UNET_block}(a)) were introduced as an improved variant of RNNs. The inherent memory cells and gating mechanisms allow LSTM models to learn and preserve relevant information, enabling them to effectively capture temporal patterns and intricate relationships between consecutive ECG samples. This ability is essential in reconstructing missing ECG leads and producing accurate representations of the underlying cardiac activity. 


\subsection{U-shaped Network (UNet)}

UNet models are a type of deep learning architecture that have been widely used for medical image analysis and segmentation tasks. 
The architecture of Unet models allows for the integration of information at different resolutions, enabling multiscale analysis (Figure \ref{LSTM_UNET_block}(b)). UNet models consist of a contracting path (encoder) and an expanding path (decoder), with skip connections that connect corresponding levels between these paths. The contracting path in the architecture is responsible for capturing context and extracting features from the input data. It gradually reduces the spatial resolution of the input while increasing the number of feature channels. On the other hand, the expanding path uses the information from the contracting path to reconstruct the output at a higher resolution. The skip connections enable the flow of information between corresponding levels in the contracting and expanding paths, allowing for the integration of multiscale information. This multiscale capability of UNet allows the model to consider both local and global context within the ECG signal, enhancing its ability to accurately reconstruct missing leads.  


\subsection{Proposed workflow}

As shown in Figure \ref{LSTM_UNET_block}(c), each input ECG record was preprocessed, i.e., resampled at 1000 Hz, baseline corrected, and normalized to the range [-1,1]. The reconstruction algorithm was trained with first 5-seconds of record and rest of the record was used for testing. Within the training set, training and a 5-fold cross-validation were carried out to validate the ML model at hand. Specifically, we performed a grid search to obtain the optimal hyperparameters for the LSTM-UNet model, with the objective of maximizing the mean coefficient of determination ($R^2$) value. 

The LSTM-UNet model was implemented using the Keras deep learning library with TensorFlow backend. The model consisted of three LSTM layers with 256, 128, and 64 units, respectively, followed by a UNet layer. Each LSTM layer utilized the rectified linear unit (ReLU) activation function to introduce non-linearity in the model. The model was trained using the Adam optimizer with a learning rate of 0.001 and a batch size of 100. To prevent overfitting, dropout with a rate of 0.2 was applied after each LSTM layer. Additionally, L2 regularization with a weight decay of 0.001 was applied on the LSTM layers to improve the model's generalization. 
In the encoder side of the UNet, we incorporated three convolution blocks, each with (64, 128, 256) filters, to extract and encode the spatial features of the ECG signal. Each convolution block was followed by a max pooling layer and leaky ReLU activation function. For the decoder side, the UNet utilized (256, 128, 64) filters along with upsampling to restore the spatial resolution of the feature maps. This design ensures accurate reconstruction of the original 12-Lead ECG signal from the compressed representation obtained from the encoder side. Finally, we applied a fully connected layer to reconstruct the 12-Lead ECG signals. The model was trained for 400 epochs, and early stopping was used to halt training if there was no improvement in performance.
\subsection{Performance metrics}

In this study, we employed the coefficient of determination ($R^2$), correlation coefficient ($r_x$), and Normalized Dynamic Time Warping ($NDTW$) as performance evaluation metrics for assessing the accuracy of ECG signal reconstruction \cite{ndtw}. 
Specifically, considering each record for a subject indexed $i$ ($i=1,2,...,n$), we denoted the original signal by $x_i$ and the reconstructed signal by $y_i$, and computed $R^2 = 1-\frac{\Sigma_{i=1}^n {(x_i-\bar{y})}^2}{\Sigma_{i=1}^n {(y_i-\bar{y})}^2}$, $r_x = \frac{\Sigma_{i=1}^n {(x_i-\bar{x})} {(y_i-\bar{y})}}{\sqrt{\Sigma_{i=1}^n {(x_i-\bar{x})}^2} \sqrt{\Sigma_{i=1}^n {(y_i-\bar{y})}^2}}$ and $NDTW = \frac{dtw(x,y)}{\sqrt{\Sigma_{i=1}^n {(x_i-y_i)}^2}}$, where $dtw(x,y) = {min(\Sigma_{i=1}^n {(x_i-{y_{\sigma(i)}})}^2})$ and $\sigma$ represents the alignment path \cite{ndtw}. Ideally,  $R^2 = 1$ for perfect reconstruction, $r_x = 1$ for strong positive correlation and $NDTW = 0$ for a perfect match between the signals. Conversely, lower values of $R^2$, $r_x$, and higher values of $NDTW$ indicate weaker reconstruction performance, suggesting the need for improvements in the ECG signal reconstruction methods.


\section{Results and Discussion}

Based on three leads, namely, I, II and V2, we reconstruct using the proposed LSTM-UNet algorithm nine additional leads, namely, III, aVR, aVL, aVF, V1, V3, V4, V5, V6. First, we picked two records from PTBDB, corresponding to a representative healthy and a diseased subject, and visually compared the reconstructed waveforms with already available ones in Figure \ref{HC_UH}.
In both cases, the match appears visually satisfactory, although the diseased record appears to return a slightly reduced performance.
Next we perform statistical analysis within subject groups associated with various heart conditions, namely, healthy control (HC),  bundle branch block (BB), hypertrophy, cardiomyopathy, and heart failure (HY), myocardial infarction (MI),  valvular myocarditis and other miscellaneous (VA), and patients with no diagnostic data (ND).

Table \ref{ptbdb_results} presents detailed results for the HC and diseased groups in the PTBDB. The LSTM-Unet model consistently exhibited the highest mean $R^2$ values, mean correlation coefficient $r_x$, and lowest mean $NDTW$ values, indicates strong agreement between the original and reconstructed ECG signals and proving superior to the reference LT and LSTM methods. While all three approaches demonstrated satisfactory performance (Mean $R^2$ value for LT=94.25\%, LSTM=96.82\%, LSTM-UNet=97.05\%) for the healthy group, the performance levels were slightly lower in diseased categories. Overall, LSTM-UNet achieved a mean $R^2$ value of 94.37\%, which was 0.79\% higher than LSTM and 2.73\% higher than LT. Analogous information is furnished in Table \ref{incart_results} for records from INCARTDB, where subgroups were not considered. Compared to PTBDB, performances were somewhat depressed for INCARTDB, possibly due to the latter's noisy nature and the presence of a wide variety of cardiac conditions. However, performance improvement appears more significant. Overall, the proposed LSTM-UNet achieved a mean $R^2$ value of 93.91\%, which was 1.78\% higher than reference LSTM and 12.17\% higher than reference LT. 

Our method, while promising, has significant room for improvement. For example, an expanded analysis should include records representing heart conditions and demographics beyond those in PTBDB and INCARTDB. Further, more subgroups within the diseased category could be more insightful. In addition, metrics beyond $R^2$, $r_x$, and $NDTW$ that are clinically relevant should be considered. Finally, beyond the current personalized ECG reconstruction algorithms, one must develop generalized algorithm that can be deployed for public healthcare centers and other locations.


\section{Conclusion}

Combining the strengths of LSTM (temporal) and UNet (spatial), the proposed method has been accurate in reconstructing 12-lead ECG from 3-lead system by effectively capturing spatio-temporal features. This potentially economizes personalized home-based cardiac monitoring systems and makes those more portable.

\end{document}